\documentclass{article}

\usepackage{bbm}
\usepackage{graphicx}
\usepackage{epstopdf}
\usepackage{listings}
\usepackage{booktabs}
\usepackage{subfigure}
\usepackage{siunitx}
\usepackage{amssymb}
\usepackage{lineno,hyperref}
\usepackage{nicefrac}
\usepackage{microtype}
\usepackage{amsmath}

\title{Characterization of Vehicle Behavior with Information Theory}

\author{Andre L. L. Aquino$^{1}$, Tamer S. G. Cavalcante$^{1}$, Eliana S. Almeida$^{1}$\\ Alejandro C. Frery$^{1}$, Osvaldo A. Rosso$^{2,3}$\\
\\
\begin{tabular}{c}
$^{1}$ IC-UFAL, Maceio, AL, Brazil \\
\url{{alla.lins, tamersgc, eliana.almeida, acfrery}@gmail.com} \\
\\
$^{2}$ IF-UFAL, Maceio, AL, Brazil
\\
$^{3}$ ITBA, Buenos Aires, Argentina\\
\url{oarosso@gmail.com}
\end{tabular}
}

\begin{document}

\maketitle

\begin{abstract} 
This work proposes the use of Information Theory for the characterization of vehicles behavior through their velocities.
Three public data sets were used:
i.~\textit{Mobile Century} data set collected on Highway I-880, near Union City, California;
ii.~\textit{Borl\"ange GPS} data set collected in the Swedish city of Borl\"ange; and 
iii.~\textit{Beijing taxicabs} data set collected in Beijing, China,
where each vehicle speed is stored as a time series.
The Bandt-Pompe methodology combined with the Complexity-Entropy plane were used to identify different regimes and behaviors.
The global velocity is compatible with a correlated noise with $f^{-k}$ Power Spectrum with $k \geq 0$.
With this we identify traffic behaviors as, for instance, random velocities ($k \simeq 0$) when there is congestion, and more correlated velocities ($k \simeq 3$) in the presence of free traffic flow.
\end{abstract}

\section{Introduction}
\label{sec:intro}
 
Vehicular Ad-hoc Networks (VANETs) are mobile ad-hoc networks where the vehicles interact among them or with a planned road infrastructure. 
In both cases, the vehicles are able to send, receive or redirect data~\cite{Hartenstein2008}.
The analysis of such networks requires understanding the variation of individual vehicles velocities~\cite{Liao2012}, among other features.
The main motivation to study the behavior of vehicles velocities is the numerous applications that use this information to propose, for instance, new prediction models~\cite{Pappalardo2013}, traffic simulators, roads conditions estimators, and real-time predictors of vehicles flow.

Characterizing velocity variability is a challenging task due to its dynamic nature.
Most available models introduce random noise in their formulation~\cite{Jabari2012}.
We will provide a better suited description, based on real-world observations, that may be used to improve both descriptions and forecasts.

The problem addressed in this work can be stated as follows:
\begin{quote}
``\textit{What can be inferred about the global behavior of vehicles through the mere observation of their velocities.}''
\end{quote}
This problem can be further assessed by answering the following questions:
(i)~Are the dynamics governing vehicles velocities stochastic, or are they deterministic chaos? 
(ii)~Are a few vehicles velocities able to characterize the traffic globally?
The latter question is related to the important problem of sampling, while the former deals with the ability of making inferences using the data at hand.

In this work, velocities of each vehicle were analyzed as a time series, and used to characterize the global velocity behavior. 
Two Information Theory quantifiers were used: the Shannon Entropy~\cite{Shannon1948} and the Statistical Complexity~\cite{LopezRuiz1995,Lamberti2004}. 
Each time series is mapped into a point in the causality Complexity-Entropy plane evaluated using the Bandt-Pompe methodology \cite{Bandt2002}.
Entropy is the system degree of ``disorder'', and the Statistical Complexity assesses the presence of structure in a process. 
This mapping allows to distinguish between determinism and randomness of vehicle velocities~\cite{Rosso2013,Rosso2007a}. 

With this methodology we verify that the global behavior of velocities is compatible with correlated noise $f^{-k}$ power spectrum, 
with $k > 0$~\cite{Rosso2007a}, and that the greater traffic congestions present uncorrelated velocities, i.e., the velocities present a random behavior.
On the order hand, free-flowing traffic presents more correlated velocities.

Three public data sets with different characteristics were used in this study: \textit{Mobile Century} data set collected on Highway I-880, near Union City, California~\cite{Herrera2010}. 
\textit{Borl\"ange GPS} data set collected in the Swedish city of Borl\"ange~\cite{Frejinger2007}. 
\textit{Beijing taxicabs} data set collected in Beijing, China~\cite{Zhu2013}. 

It is important to highlight that we use only the individual vehicle velocities with no additional assumptions or communication requirements.
Other works, as in Ref.~\cite{SnMf_2013}, characterize traffic behavior using data shared among vehicles.
Based on this communication, vehicles are described as forming a large network and, subsequently, it topology is analyzed.
This aspect was not considered in our work because we are interested in using the simplest data available in a VANET, i.e., the vehicle velocities. 
Other communication-dependent topology analyses will be considered in future works.

Among the contributions of this work we highlight: 
the use of Information Theory concepts to analyse vehicular networks data; 
the proposition of a methodology to characterize vehicle velocities from real data sets; and 
the characterization of vehicle velocities behavior through the causality Complexity-Entropy plane.

This work is organized as following: 
Section~\ref{sec:relacionados} presents the related work.
Section~\ref{sec:methods} explains the Information Theory quantifiers.
Section~\ref{sec:data} discusses the VANETs data set used.
Section~\ref{sec:stochastic_data} shows the stochastic $f^{-k}$ power spectrum data.
Section~\ref{sec:results} relates the results obtained.
Finally, Section~\ref{sec:conclusions} concludes the manuscript.
 
\section{Related work}
\label{sec:relacionados}

Many works employed the Complexity-Entropy plane for the characterization of time series in a diversity of applications: stock market evolution~\cite{Zanin2012}, brain behavior~\cite{Zanin2012}, climate change~\cite{Carpi2012}, and vehicles velocities variation~\cite{Liao2012}. 

Liao et al.~\cite{Liao2012} analyzed the traffic congestion index (TCI) of Beijing which is defined by the Beijing Municipal Commission.
The TCI is computed using information gathered by specialized devices (in vehicles, traffic lights etc.), all around the city.
This index indicates the traffic state as congested or not.
Using the Bandt-Pompe methodology, the authors showed that the Complexity-Entropy plane is the best way to classify the levels of traffic congestion using the TCI as reference, when compared to other methods: $R/S$ analysis (\emph{Rescaled Range})~\cite{Hurst1951} and DFA method (\emph{Detrended Fluctuation Analysis})~\cite{Peng1995}.
The former is a statistical method used to evaluate the nature and the magnitude of the variability of the data over time, 
while the latter determines the behavior of data scale considering some trends, regardless their origin or shape.
Differently from our approach, Ref.~\cite{Liao2012} does not attempt at characterizing the underlying dynamics of the process.

Liao et al. present the position of each velocity group in the plane every day of the week at two moments (Morning and Evening), and correlate this information with that provided by the TCI.
In our work, a global characterization is presented based solely on data measured from selected vehicles.
We conclude that vehicle velocities are compatible with correlated noise 
with $f^{-k}$ Power Spectrum, $k \geq 0$, so we can use this result to improve the vehicle traffic studies, for instance, vehicle behavior characterization or traffic simulation tools.

Other approaches used to characterize the vehicles velocities are:
\begin{itemize}
\item Shang et al.~\cite{Shang2006} use the fractal dimension to analyze vehicles velocities in the \textit{Beijing Yuquanying highway} data set collected by Highway Performance Measurement Project, in Beijing, China. 
The data were collected by using sensors on the roads which sampled the velocities in intervals of \SI{20}{\second}. 
Results showed that the traffic, in the scenarios considered, have multifractal characteristics and the degree of fractality increases proportionally to traffic congestion.  
Moreover, the local H\"older exponent (or roughness)~\cite{Daoudi1998} was used to predict heavy traffic. 
This exponent can be used to measure the local rate of fractality.

\item Pappalardo et al.~\cite{Pappalardo2013} compared human and vehicle mobility patterns, trying to apply models of the former to the latter.
The results were twofold: 
i.~Known models of human mobility can be refined to treat the vehicles mobility; and 
ii.~They proved that the GPS data are an adequate representation of vehicles mobility.
The data set used was collected in Italy with information about $150,000$ vehicles and approximately $10$ million trajectories.

\item Tang et al. \cite{Tang2013} characterized time series of vehicle velocities using Complex Networks.  
They transformed the time series into a new series based on the reconstruction of the phase space.
Then, complex networks of traffic flow were built and analyzed considering the degree distribution, density, and clustering coefficient.
The results detected that the networks present communities structures, i.e., the nodes of the networks are grouped so that each group is highly connected.
The data set used was collected in the city of Harbin, China, during one complete day with sampling at \SI{2}{\minute}, resulting in $720$ samples.
\end{itemize}

We use data sets where the vehicles themselves register their GPS location allowing the extraction of velocities information.
On the one hand, data collected by GPS are more susceptible to errors, require a more accurate treatment and, hence, more robust methods for their analysis.
On the other hand, they are the least demanding in terms of new infrastructure requirements.
We use Information Theory concepts, specifically, the Shannon Entropy, the Statistical Complexity evaluated using Bandt-Pompe symbolization method and the causal Complexity-Entropy plane to characterize the vehicles velocities.
With this, we provide a more detailed analysis of the global vehicles velocities dynamics by identifying if the velocities have chaotic or stochastic behavior.
This allows us to compare the velocities behavior with colored noises, just observing their location on the causal Complexity-Entropy plane.

\section{Time series and Information Theory based quantifiers}
\label{sec:methods}

\subsection{Shannon entropy and statistical complexity}

Information Theory quantifiers are measures able to characterize properties of the probability distribution function (PDF) associated with an observable or measurable quantity.
Entropy, regarded as a measure of uncertainty, is the most paradigmatic example~\cite{Shannon1948}.
Kolmogorov and Sinai converted Shannon's Information Theory into a powerful tool for the study of dynamical systems~\cite{Kolmogorov1958}.
 
Let $x$ be a discrete random variable with $N<\infty$ possible values ${\mathcal X} = \{ x_j : j= 1, \dots, N \}$ whose distribution is characterized by the probability function $P = \{p_i : i = 1, \dots, N\}$.
The widely known Shannon logarithmic information measure~\cite{Shannon1948} is
\begin{equation}
{\mathcal S}[P] =- \sum_{i=1}^{N} p_i \ln{p_i},
\label{eq:Shannon-entropy}
\end{equation}
where $a \ln{a} = 0$ if $a=0$ by definition.
The Shannon entropy is related to the information associated with the physical process 
described by $P$. 
If ${\mathcal S} [P] = 0$ the knowledge about the underlying process described by $P$ is maximal and the possible outcomes can be predicted with complete certainty. 
On the other hand, our knowledge is minimal for a uniform distribution 
$P_e = \{ p_i = 1/N, \forall i\}$ since every outcome exhibits the same probability 
of occurrence. 
It is well known, however, that the degree of structure present in a process is not 
quantified by randomness measures and, consequently, measures of statistical or 
structural complexity are necessary for a better understanding of chaotic time series~\cite{Feldman1998}.

There is no universally accepted definition of complexity. 
Intuitively, complexity should be related to the amount of structure or the number 
of patterns present in a system. 
One would like to have some functional ${\mathcal C}[P]$ able to capture the ``structuredness" in the same way as Shannon's entropy \cite{Shannon1948} captures randomness.

The perfect crystal and the isolated ideal gas are two typical examples of systems with minimum and maximum entropy, respectively. 
However, they are also examples of simple models and therefore of systems with zero 
complexity, as the structure of the perfect crystal is completely described by minimal
information (i.e., distances and symmetries that define the elementary cell) and the 
probability distribution for the accessible states is centered around a prevailing 
state of perfect symmetry. 
On the other hand, all the accessible states of the ideal gas occur with the same 
probability and can be described by a ``simple" uniform distribution. 
According to L\'opez-Ruiz et al.~\cite{LopezRuiz1995}, and using a tautology, an object, 
a procedure, or system is said to be complex when it does not exhibit patterns regarded 
as simple. 
It follows that a suitable complexity measure should vanish both for completely ordered 
and for completely random systems and cannot only rely on the concept of information 
(which is maximal and minimal for the above mentioned systems). 
  
A suitable measure of complexity can not be made in terms of just ``disorder" or 
``information". 
It seems reasonable to propose a measure of ``statistical complexity"  by adopting 
some kind of distance to a reference probability distribution, in particular to the 
uniform distribution $P_e$  \cite{LopezRuiz1995,Lamberti2004}. 
This motivates introducing, as a special distance-form, the so-called 
``disequilibrium-distance" ${\mathcal Q}[P,P_e]$.
In this respect, Rosso et al. \cite{Lamberti2004} introduced an effective 
statistical complexity measure that is able to detect essential details of important dynamics. 
This statistical complexity measure is defined following the functional product proposed by 
L\'opez-Ruiz et al. \cite{LopezRuiz1995}:
\begin{equation}
{\mathcal C}_{JS}[P] =  {\mathcal H}_S[P] \, {\mathcal Q}_{JS}[P,P_e] .
\label{eq:complexity}
\end{equation}
where
\begin{equation}
{\mathcal H}_S[P] = {\mathcal S}[P] / \mathcal S^{\max}
\label{eq:normalized-shannon}
\end{equation}
is the normalized Shannon entropy ${\mathcal H}_S \in [0,1]$ with 
$\mathcal S^{\max} = {\mathcal S}[P_e] = \ln N$,
and the disequilibrium ${\mathcal Q}_{JS}$ is defined in terms of the extensive 
(in the thermodynamical sense) Jensen–-Shannon divergence. 
Namely, 
\begin{align}
{\mathcal Q}_{JS}[P,P_e] &= {Q}_0 \, {\mathcal J}_{S}[P,P_e]  \nonumber \\
&= {Q}_0  \left\{ \mathcal S\left[ {P + P_e} \over {2} \right] - {{1}\over{2}} \mathcal S[P] - {{1}\over{2}} \mathcal S[P_e]  \right\} ,
\label{eq:disequilibrium}
\end{align}
and $Q_0$ is a normalization constant equal to the inverse of the maximum
possible value of ${\mathcal J}_{S}[P,P_e]$ so that ${\mathcal Q}_{JS} \in [0,1]$.
$Q_0$ is obtained when one of the probabilities of $P$ is equal to one and
the remaining are equal to zero. 

The Jensen-Shannon divergence quantifies the difference between 
probability distributions, and is well-suited to compare the symbol composition 
between different sequences~\cite{Grosse2002}. 
The statistical complexity of a system is null in the
opposite extreme situations of perfect knowledge (perfect crystal), and maximal 
randomness (ideal gas), whereas a wide range of possible degrees of physical structure 
does exist between these extreme configurations.

The statistical complexity in Eq.~(\ref{eq:complexity}) is not a trivial 
function of the entropy because it measures the interplay between the information
stored by the system and the distance from equipartition (measure of a probabilistic 
hierarchy between the observed parts) of the probability distribution of its accessible 
states~\cite{LopezRuiz1995}. 
Furthermore, a range of possible statistical complexity values does exist for 
any non-null ${\mathcal H}_S$ value~\cite{Martin2006}, that is 
${\mathcal C}_{JS}^{\min} \leq {\mathcal H}_S \leq {\mathcal C}_{JS}^{\max}$, 
meaning that additional information related to the dependence structure between the 
components of the system and the emergence of nontrivial  collective behavior is provided 
by evaluating the statistical complexity.

Moreover, it should be noted that statistical complexity fulfills two additional properties 
required for a suitable definition of complexity~\cite{LopezRuiz2011}: 
(1)~the quantifier must be measurable in different physical systems, and 
(2)~it should allow for physical interpretation and comparison between two measurements.
Indeed, the definition of complexity in Eq.~(\ref{eq:complexity}) also depends on the scale. 
For a given system at each scale of observation, a new set of accessible states appears with
its corresponding probability distribution, so that complexity changes and therefore different 
values for ${\mathcal H}_S$ and ${\mathcal C}_{JS}$ are obtained. 
 
\subsection{Bandt-Pompe symbolization method}

The evaluation of ${\mathcal H}_S$ and ${\mathcal C}_{JS}$ requires the definition 
of a probability distribution $P$ associated with the time series.
Bandt and Pompe~\cite{Bandt2002} introduced a simple method to define this probability distribution
taking into account the time causality of the process.
 This approach is based on the symbol sequences that arise naturally from the time series, replacing 
the observed series with a sequence of ranks. 

Given a time series ${\mathcal X}(t) = \{x_t:  t = 1, \dots, M \}$, an embedding
dimension $D \geq 2$ ($D \in {\mathbb N}$), and an embedding delay time $\tau\in {\mathbb N}$,
the ordinal pattern of order  $D$ (pattern length) generated by
\begin{equation}
(s)\mapsto \left(x_{s-(D-1)\tau}, x_{s-(D-2)\tau}, \dots, x_{s-\tau}, x_{s}\right) \ ,
\label{vectores}
\end{equation}
is considered. 
For each time instant $s$, we assign a $D$ dimensional vector that results from the evaluation of the
time series at times $s - (D - 1)\tau ,\dots , s - \tau, s$. 
Clearly, the higher the value of $D$, the more information about the past is incorporated into the ensuing vectors. 
By the ordinal pattern of order $D$ related to the time instant $s$ we mean the permutation 
$\pi = \{r_0, r_1, \dots, r_{D-1}\}$ of $\{0, 1, \dots , D - 1\}$
defined by
\begin{equation}
x_{s-r_{D-1}\tau}\le x_{s-r_{D-2}\tau} \le \cdots \le x_{s-r_{1}\tau} \le x_{s-r_0 \tau} .
\label{permuta}
\end{equation}
In this way the vector defined by Eq.~(\ref{vectores})  is converted into a unique symbol $\pi$. 
In order to get a unique result we set $r_i <r_{i-1}$ if $x_{s-r_{i}}=x_{s-r_{i-1}}$.
Equal values have probability zero if the values of $x_t$ follow a continuous distribution.

Thus, for all the $D!$ possible permutations $\pi$ of order $D$, their associated relative 
frequencies can be naturally computed by the number of times this particular order sequence 
is found in the time series divided by the total number of sequences.
The probability distribution $P \equiv \{p(\pi)\}$ is defined by
\begin{equation}
p(\pi)= \frac{\# \{s \text{ of type }\pi : s\leq M-(D-1) \tau  \}}{M-(D-1) \tau},
\label{frequ}
\end{equation}
where $\#$ is the cardinality of the set.
This probability distribution is linked to the sequences of ranks resulting from the comparison of
consecutive ($\tau = 1$) or non-consecutive ($\tau > 1$) points, allowing for the empirical reconstruction
of the underlying phase space~\cite{Bandt2002}. 

It is worth noting that the method is rank-based, and the ordinal pattern probability distribution 
is invariant with respect to monotonic transformations. 
Thus, nonlinear drifts or scaling artificially introduced by a measurement device do not modify 
the quantifier estimations. 
This property is highly desired for the analysis of experimental data and natural time series 
analysis~\cite{Carpi2012}. 
Nevertheless, the rank-based description unavoidably results in loss of information, which is however 
common to every nonparametric rank based statistical method.
Additional advantages of the method reside in 
{\it i)\/} its simplicity (few parameters are needed: the pattern length/embedding dimension $D$,
and the embedding delay $\tau$,  and 
{\it ii)\/} the extremely fast nature of the calculation process~\cite{Keller2007}. 

The Bandt-Pompe methodology can be applied not only to time series representative of low dimensional
dynamical systems but also to any type of time series (regular, chaotic, noisy, experimentally
obtained or artificially generated). 
In fact, the existence of an attractor in the $D$-dimensional phase space is not assumed. 
The only condition for the applicability of the Bandt-Pompe methodology is a very weak stationary 
assumption, that is, for $k = D$, the probability for $x_t < x_{t+k}$ should not depend on $t$~\cite{Bandt2002}.

The probability distribution $P$ is obtained once we fix the embedding dimension $D$ and the embedding 
delay time $\tau$. 
The former parameter plays an important role for the evaluation of the appropriate probability 
distribution, since $D$ determines the number of accessible states $D!$, therefore the length $N$ of the time series must satisfy the condition $N \gg D!$ in order to obtain reliable statistics. 
Bandt and Pompe \cite{Bandt2002} considered an embedding delay $\tau = 1$. 
Nevertheless, other values of $\tau$ might provide additional information. Zunino et al.~\cite{Zunino2012b} showed that this parameter is strongly related, when it is relevant, to the intrinsic time scales of the system under analysis.

In this work we evaluate the normalized Shannon entropy ${\mathcal H}_S$ and the statistical
complexity ${\mathcal C}_{JS}$ using the permutation probability distribution $P \equiv \{ p(\pi)\}$ 
(the PDF-Bandt-Pompe), so that the former quantifier is called {\it permutation entropy\/} 
and the latter {\it permutation statistical complexity.\/}

\subsection{The causality entropy-complexity plane}

In statistical mechanics one is often interested in isolated systems characterized by an initial, 
arbitrary, and discrete probability distribution. 
Evolution towards equilibrium is to be described, as the overriding goal. 
At equilibrium, we can think, without loss of generality, that this state is given by the uniform 
distribution $P_e$. 
The temporal evolution of the Shannon entropy and the statistical complexity measure can 
be analyzed using the two-dimensional diagram of  ${\mathcal H}_S$ and ${\mathcal C}_{JS}$ versus 
time $t$. 
However, it is well known that the second law of thermodynamics states that 
entropy grows monotonically with time ($d{\mathcal H}_S/dt \geq 0$) in isolated systems~\cite{Plastino1996}. 
This implies that  ${\mathcal H}_S$ can be regarded to as an arrow of time, so that an equivalent way 
of studying the temporal evolution of the statistical complexity is through
the analysis of ${\mathcal C}_{JS}$ versus ${\mathcal H}_S$. 
In this way, the normalized entropy-axis replaces the time-axis. 
Furthermore, as we mention previously, it has been shown that for a given value of ${\mathcal H}_S$, 
the range of possible statistical complexity values varies between a minimum ${\mathcal C}_{JS}^{\min}$ 
and a maximum  ${\mathcal C}_{JS}^{\max}$, restricting the possible values 
of the  statistical complexity in this plane~\cite{Martin2006}. 

The {\it entropy--complexity causality plane\/} is defined as the two-dimensional
diagram obtained by plotting permutation statistical complexity (vertical axis) versus the permutation
entropy (horizontal axis) for a given system~\cite{Rosso2007a}.
The term ``causality'' reminds the fact that temporal correlations between successive samples are
taken into account through the PDF-Bandt-Pompe used to estimate both Information Theory quantifiers. 

This diagnostic tool was shown to be particularly efficient at distinguishing between the deterministic 
chaotic and stochastic nature of time series, since the permutation quantifiers have distinctive
behaviors for different types of dynamics. 
According to the findings obtained by Rosso et al.~\cite{Rosso2007a}, chaotic maps have intermediate 
${\mathcal H}_S$ values, while ${\mathcal C}_{JS}$ reaches larger values, close to those of the limit
${\mathcal C}_{JS}^{\max}$. 
For regular processes, both quantifiers have small values, close to ${\mathcal H}_S = 0$ and
${\mathcal C}_{JS} = 0$. 
Finally, totally uncorrelated stochastic processes are in the planar location associated with 
${\mathcal H}_S \approx 1$ and ${\mathcal C}_{JS} \approx 0$, respectively.

Rosso et al.~\cite{Rosso2007B} also found that $f^{-k}$ power spectrum  correlated stochastic processes with 
$0 < k \leq 3$ are characterized by intermediate permutation entropy and intermediate statistical complexity 
values between ${\mathcal C}_{JS}^{\min}$ and ${\mathcal C}_{JS}^{\max}$.
Similar planar localization were also found for fractional Brownian motion (fBm) and fractional 
Gaussian noise (fGn) dynamics~\cite{Rosso2013,Rosso2007a,Rosso2007B}. 

The planar localization for chaotic noise dynamics was also studied by Rosso et al. \cite{Rosso2012A,Rosso2012B}.
It was shown that the localization in the entropy--complexity causality plane of the logistic map (fully developed chaos) contaminated with uncorrelated or correlated additive noise of different amplitude, changes from the original location towards the location of pure noise as the noise amplitude increases, but following quite close to the curve
of maximum complexity ${\mathcal C}_{JS}^{\max)}$. 
This allows an easy identification of chaotic--noise behavior.

\subsection{On the time sample and causality ${\mathcal H}\times {\mathcal C}$-plane}
\label{sec:time_vs_plane}
 
Dynamical systems may present different types of behavior depending on the time scale. 
Therefore, the scale and those system parameters values at which any study is made must 
be specified. 
The study of dynamical systems throughout the use of observed data, usually time series, 
is quite frequent.

On the one hand, when the independent variable, usually the time, is discrete, the dynamical  
process under study is an {\it iterated map\/} that generates the corresponding time series. 
In this case, each time step is the natural sampling time.
For continuous time, dynamical processes characterized by {\it differential equations\/} 
produce data in a continuous fashion, a type of process also known as a dynamical flow.  
Dynamical flows are often  the result of a system described by ordinary differential equations 
(ODEs), and their outcome can be obtained by integrating these ODEs.
Such resolution must be performed at specified integration times and, as a consequence,  
there is a \textit{natural} or \textit{ideal} sampling time at which the type of dynamics 
under study can be clearly identified.

The {\it Causal Statistical Complexity\/}, ${\mathcal C}_{JS}$, is obtained by first computing 
the Band \& Pompe  probability distribution function (PDF) associated to the time 
series~\cite{Bandt2002}, and then computing Information Theory quantifiers: entropy, ${\mathcal H}_S$ 
and disequilibrium ${\mathcal Q}$~\cite{LopezRuiz1995,Lamberti2004}. 
This PDF is not a dynamical system invariant, but under mild conditions it exhibits little variation 
with respect to the total length $N$ of the time series. 
The condition for this stability is that $N \gg D!$, where $D$ is the pattern length under analysis. 
Besides the Causal Statistical Complexity, the {\it Normalized Permutation Entropy\/} (Normalized 
Shannon Entropy), ${\mathcal H}_S$, is another useful quantifier based on the Bandt \& Pompe PDF. 
With these elements, Rosso and collaborators introduced the {\it causal Entropy--Complexity plane\/} 
${\mathcal H} \times {\mathcal C}$~\cite{Rosso2007a}. 
This plane is bounded between the curves of minimal and maximal complexity for each value of entropy
~\cite{Martin2006}. 
These curves are solely defined by the dimension of the PDF.

Again, a dynamic produces a time series that is turned into a PDF that, in turn, becomes a point in 
the ${\mathcal H} \times {\mathcal C}$-plane. 
The position of such point is related to the type of (unobserved) dynamics under study. 
The relationship between this localization and the sampling time is, therefore, indirect, as the Bandt 
\& Pompe PDF is the central link between the dynamics and the point in the 
${\mathcal H} \times {\mathcal C}$-plane.

The signal with which the PDF is built may be smoother than the original one, if the sampling time is 
much lower than the optimal, and in consequence its time extension is not enough for revealing the real 
generating dynamics. 
This may lead to a characterization in the ${\mathcal H} \times {\mathcal C}$-plane suggesting a signal 
correlation much higher than the true one, e.g., a regular dynamic. 
Contrarily, if the signal is oversampled, the characterization may correspond to a signal with a smaller 
correlation (or even null correlation) than the true generating dynamics, i.e., to a noisier system. 

Rosso and collaborators analysed the problem of optimal time sampling, and proposed using the one 
that maximizes the causality statistical complexity~\cite{Zunino2012b,Demicco2012}. 
This can be easily obtained in two cases, namely when the data is obtained integrating ODEs, and when 
the original signal is oversampled. 
The last situation corresponds to experimental setups for which there is control over the measuring 
device and over the process under study.

When the analyst receives the signal already measured, the usual course of action is assuming that it 
was optimally sampled, as this is the only available information. 
At most, small changes in the sample time are allowed, but no new information can be produced by finer 
sampling times; this could produce the change to another time scale and, as a consequence, to a different 
dynamical behavior.
Experimental setups, therefore, lead to rather stable points in the causality 
${\mathcal H} \times {\mathcal C}$-plane (for a given $N$ and $D$), as they are characteristic of the available 
information about the underlying dynamics which generates the corresponding time series under analysis. 
Bearing this in mind, one is able to compare such points and infer about the unobserved dynamics through 
the information available to the analyst, even when the time series are obtained at different sampling times.

\section{VANETs data processing} 
\label{sec:data}

\subsection{Data sets description} 

\begin{figure*}[ht]
\centering
\subfigure[\emph{Mobile Century}]{\includegraphics[width=.32\textwidth]
	{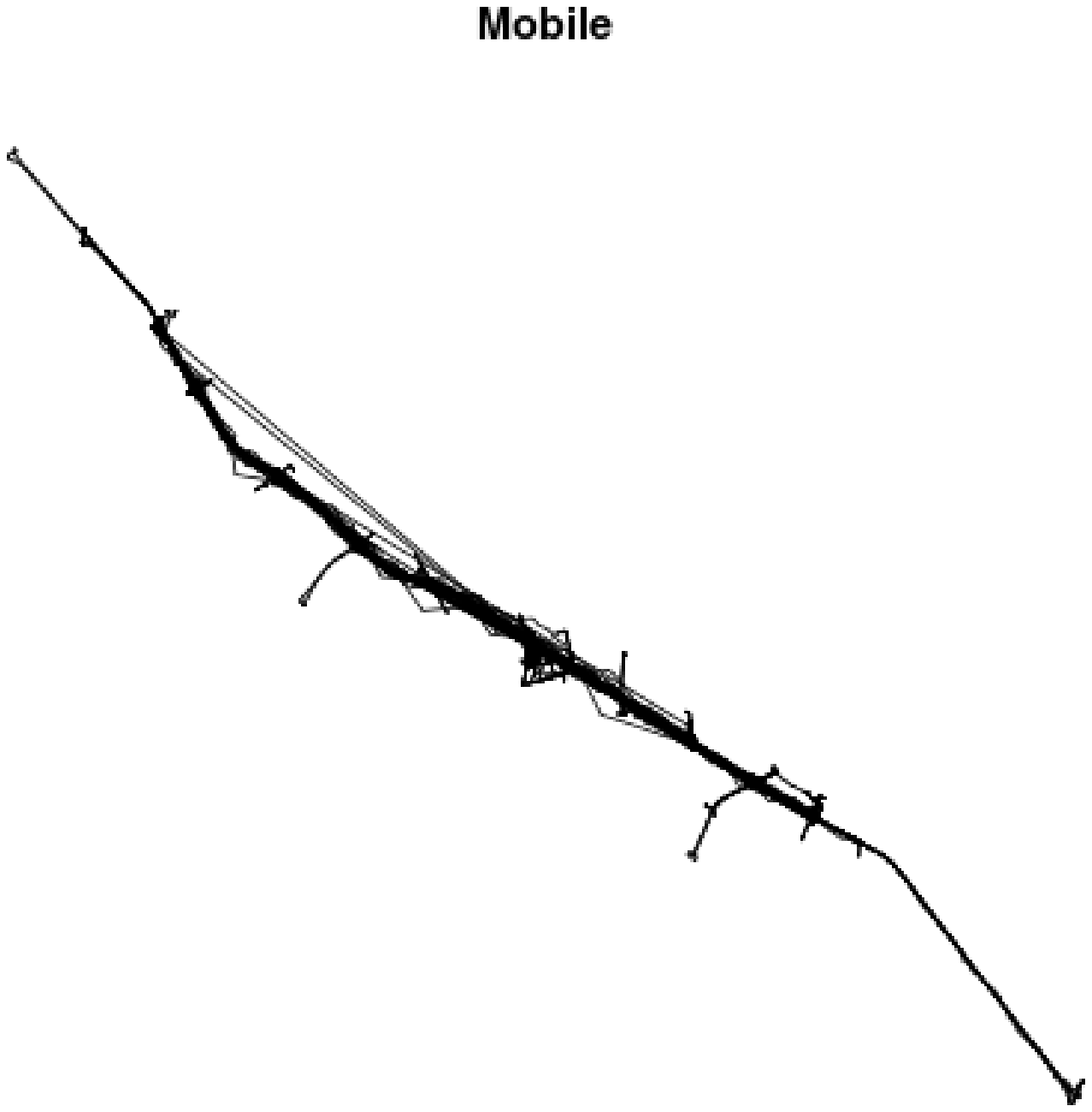}\label{fig:mapa_mobile} }
\subfigure[Borl\"ange]{\includegraphics[width=.32\textwidth]
{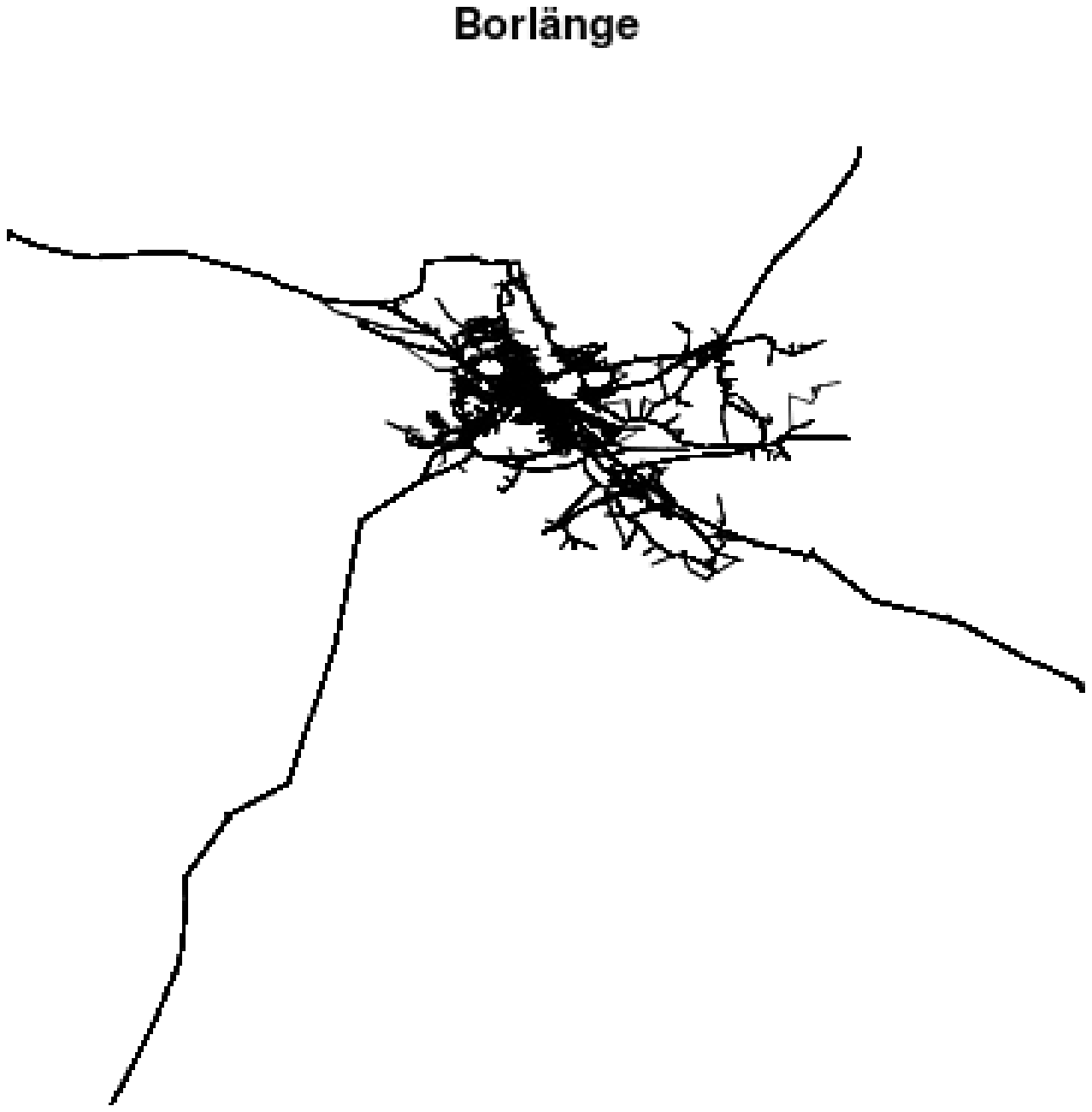}\label{fig:mapa_borlange} } 
\subfigure[Beijing]{\includegraphics[width=.32\textwidth]
{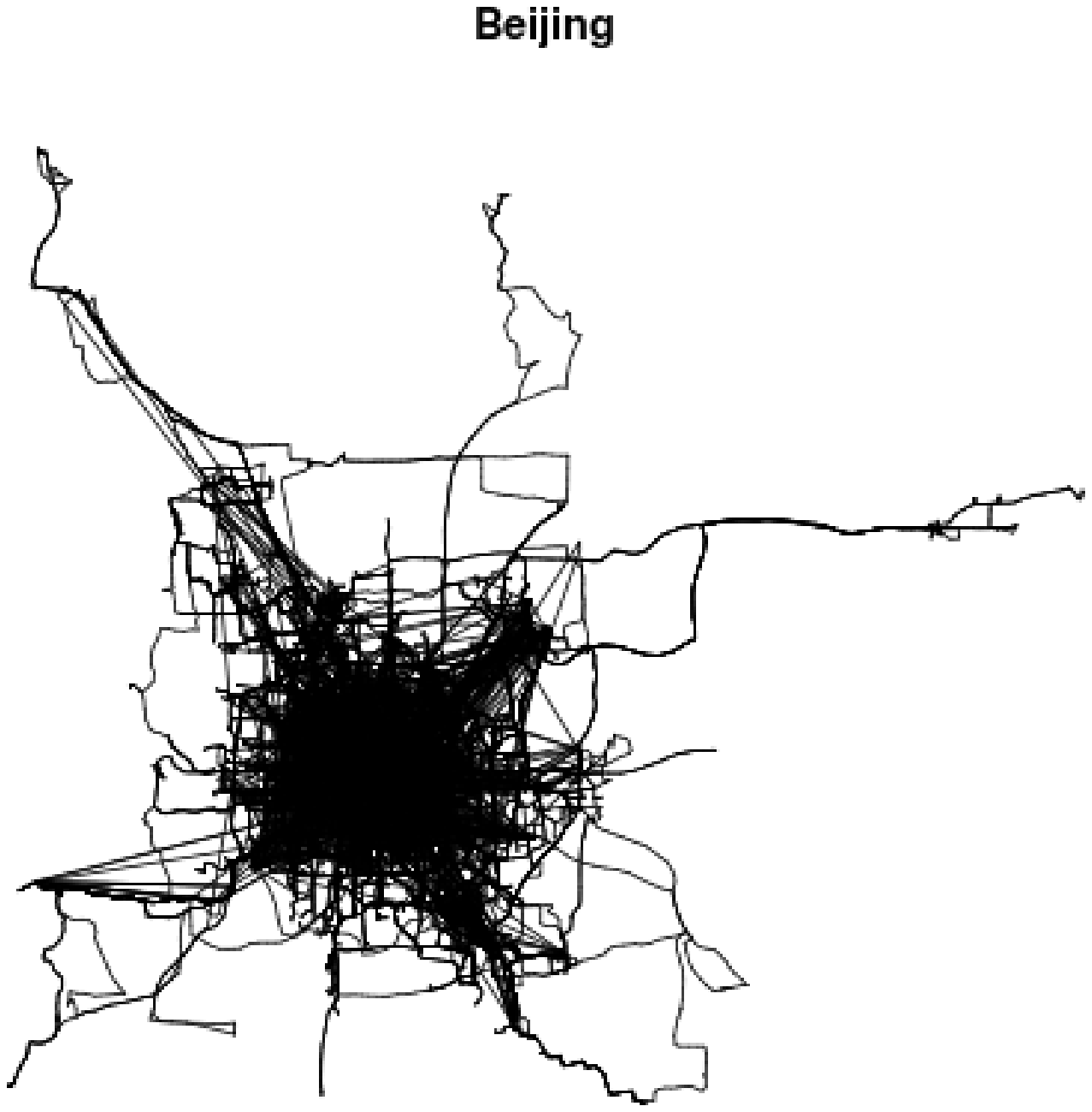}\label{fig:mapa_beijing}  }
\caption{All GPS position of each vehicle registered in each data set}
\label{fig:mapas_bases} 
\end{figure*}

As mentioned, three real data set were used to characterize the velocities behavior.
Figure~\ref{fig:mapas_bases} illustrates the road structure of each data set, as formed by the collection of all GPS positions of each registered vehicle.

\begin{itemize}
\item \textit{Mobile Century} data set collected on Highway I-880, near Union City, California, between Winton Ave.\ to the North and Stevenson Blvd.\ to the South~\cite{Herrera2010}. 
This data set (Figure~\ref{fig:mapa_mobile}) covers part of a highway with a reduced number of crossroads, and the information describes little traffic congestion.
\item \textit{Borl\"ange GPS} data set collected in the Swedish city of Borl\"ange~\cite{Frejinger2007}. 
This data set (Figure~\ref{fig:mapa_borlange}) covers a central business district with many avenues with some secondary streets, and describes a moderate occurrence of traffic congestion.
\item \textit{Beijing taxicabs} data set collected in Beijing, China~\cite{Zhu2013}.
This data set (Figure~\ref{fig:mapa_beijing}) covers a central business area with a high density of vehicles, and describes a greater occurrence of traffic congestion.
\end{itemize}

In order to use these data set in our characterization, it was necessary to extract or calculate the vehicles velocities in \si{\metre\per\second}. 
This action is described in the following Sections.

\subsubsection{\emph{Mobile Century} data set} 
 
The \emph{Mobile Century} data set contains $77$ individual GPS logs, i.e. $77$ vehicles monitored, extracted from \textit{Nokia N95} mobile devices.
They were collected in February 8, 2008, on Interstate 880, California, between 10:00~AM and 6:00~PM. 
The information available for each vehicle, collected approximately at each \SI{3}{\second}, are time, latitude, longitude and velocity in miles per hour~\cite{Herrera2010}. 

Listing~\ref{code:mobile} presents the information of a single vehicle.
The first column shows the \emph{Unix time} in milliseconds (\si{\milli\second}). 
The \emph{Unix time} is a system to describe time instants, which is defined as the number of seconds that have elapsed since January 1, 1970~\cite{Ritchie1984}.
The second and third columns present the latitude and longitude coordinates, respectively. 
The last column shows the vehicle velocity in \si{mi\per\hour}, which was later converted to \si{\meter\per\second}.

\noindent
\begin{minipage}[hbt]{1\textwidth}
\begin{scriptsize} 
\begin{lstlisting}[caption={Part of \textit{Mobile Century} data set}, label=code:mobile] 
...
1202497202837, 37.6004328519, -122.0637571325, 0.009 
1202497206836, 37.6004329358, -122.0637568810, 0.010 
1202497209837, 37.6004329358, -122.0637566295, 0.013 
1202497212837, 37.6004329358, -122.0637563781, 0.015 
1202497216826, 37.6004330196, -122.0637560428, 0.016 
1202497220826, 37.6004331872, -122.0637556237, 0.017 
...
\end{lstlisting} 
\end{scriptsize}
\end{minipage}

\subsubsection{\textit{Borl\"ange GPS} data set}
\label{subsec:borlange} 
 
The \textit{Borl\"ange GPS} data set contains 3,077 intersection connected by 7,459 roads. 
Data were collected during two years and involved 200 vehicles. 
Vehicles were equipped with a GPS and monitored in a radius of \SI{25}{\kilo\metre} from the center of the city. 
Their positions were recorded at intervals of, approximately, \SI{20}{\second}. 
Only 24 vehicles and their 420,814 GPS observations \cite{Frejinger2007} were used. 

Each entry is distributed in three different files: \texttt{mobility}, \texttt{nodes} and \texttt{nodepos}.
\texttt{mobility} registers the instants of each GPS observation,
as illustrated in Listing \ref{code:mobility}.
The first column identifies the vehicle, in this sample the vehicle is $4$. 
The second column relates the information regarding the day of observation of the vehicle, in this sample it is the first one.
Specifically to vehicle 4, the data set registered 150 more days.
The third column shows the trip number of the vehicle; in this sample, it is the second trip vehicle $4$ in the first day.
The last two columns are, respectively, the start and end time of each GPS interval observation; this will be used to estimate the velocities.

\noindent
\begin{minipage}[hbt]{1\textwidth}
\begin{scriptsize} 
\begin{lstlisting}[caption={Part of \texttt{mobility} file of \textit{Borl\"ange GPS} data set}, label=code:mobility]
...
4, 1, 2, 2000-11-10 14:24:11, 2000-11-10 14:24:19 
4, 1, 2, 2000-11-10 14:24:19, 2000-11-10 14:24:33 
4, 1, 2, 2000-11-10 14:24:33, 2000-11-10 14:24:59 
4, 1, 2, 2000-11-10 14:24:59, 2000-11-10 14:25:18 
4, 1, 2, 2000-11-10 14:25:18, 2000-11-10 14:25:23 
4, 1, 2, 2000-11-10 14:25:23, 2000-11-10 14:26:17 
4, 1, 2, 2000-11-10 14:26:17, 2000-11-10 14:26:32 
4, 1, 2, 2000-11-10 14:26:32, 2000-11-10 14:26:32 
4, 1, 2, 2000-11-10 14:26:32, 2000-11-10 14:26:36 
...
\end{lstlisting} 
\end{scriptsize} 
\end{minipage}

The \texttt{nodes} file (Listing~\ref{code:nodes}) has the same number of rows of the \texttt{mobility} one.
Each line has two values used to identify the initial and final position considering the times presented in the correspondent line in \texttt{mobility} file.

\noindent
\begin{minipage}[hbt]{1\textwidth}
\begin{scriptsize} 
\begin{lstlisting}[caption=Part of \texttt{nodes} file of \textit{Borl\"ange GPS} data set,label=code:nodes] 
...
316	1076  
1076	316  
316	792  
792 	2611 
2611	321 
321 	1823 
1823	318 
...
\end{lstlisting} 
\end{scriptsize} 
\end{minipage} 

The \texttt{nodepos} file (Listing~\ref{code:nodepos}) presents the real latitude and longitude coordinates of each identifier in the \texttt{nodes} file.
The first column shows the position identifier.
The last two columns are the latitude and longitude coordinates.

\noindent
\begin{minipage}[hbt]{1\textwidth}
\begin{scriptsize} 
\begin{lstlisting}[caption=Part of \texttt{nodepos} file of \textit{Borl\"ange GPS} data set, label=code:nodepos] 
... 
316	15.443687, 60.476045 
1076	15.445492, 60.474991 
792	15.442580, 60.475656 
2611	15.440816, 60.477419 
321	15.440701, 60.477410
1823	15.438019, 60.476698 
318	15.441260, 60.475159 
... 
\end{lstlisting} 
\end{scriptsize} 
\end{minipage} 

Combining all files (as presented in Listings~\ref{code:mobility}, \ref{code:nodes}, and~\ref{code:nodepos}), we infer that vehicle $4$ traveled from point 316 ([latitude, longitude] = [60.476045, 15.443687]) 
to 1076 ([latitude, longitude] = [60.474991, 15.445492]), leaving at \SI{14}{\hour} \SI{24}{\minute} \SI{11}{\second} in 10 November 2000 and arriving at \SI{14}{\hour} \SI{24}{\minute} \SI{19}{\second}, the same day, on its second trip, first day of recording.

This data set has not explicitly the velocity information. 
To use this information in our characterization, we calculate the average speed of each vehicle by using the time and position of each vehicles in each trip.

\subsubsection{\emph{Beijing taxicabs} data set} 

The \textit{Beijing taxicabs} data set contains routes information of $28.000$ taxis, approximately $42\% $ of all taxis in Beijing, China, at the time of the data collection.
This data set was available until December 2013 by \emph{Complex Engineered Systems Lab}\footnote{http://sensor.ee.tsinghua.edu.cn/datasets.php}.
Data collection was carried out during a month sampling, on average, every \SI{60}{\second}.
The data set is divided in 30 binary files, each for one day of collection. 

Listing~\ref{code:beijing} presents the information of all vehicles.
The first column is the vehicle identifier.
The second column presents the Coordinated Universal Time \texttt{UTC}~\cite{Nelson2001}.
The third and fourth columns contain the latitude and longitude coordinates, respectively. 
The last column is the velocity, however the unit of measure is not indicated, so it was ignored. 
We obtained the average speed of each vehicle using its time and position in each trip.

\noindent
\begin{minipage}[hbt]{1\textwidth}
\begin{scriptsize} 
\begin{lstlisting}[caption=Part of \emph{Beijing taxicabs} data set, label=code:beijing] 
...
156, 1241107200, 4000311, 11630912, 411
157, 1241107200, 3999001, 11648450, 1039
158, 1241107200, 3997923, 11644893, 771
159, 1241107200, 3993850, 11642839, 1034
160, 1241107200, 3989621, 11647153, 481
161, 1241107200, 3997600, 11641126, 370
162, 1241107200, 3992247, 11629478, 668
...
\end{lstlisting} 
\end{scriptsize} 
\end{minipage}

\subsection{Data set processing} 
 
\begin{figure}[h] 
\centering 
\includegraphics[width=1\linewidth]{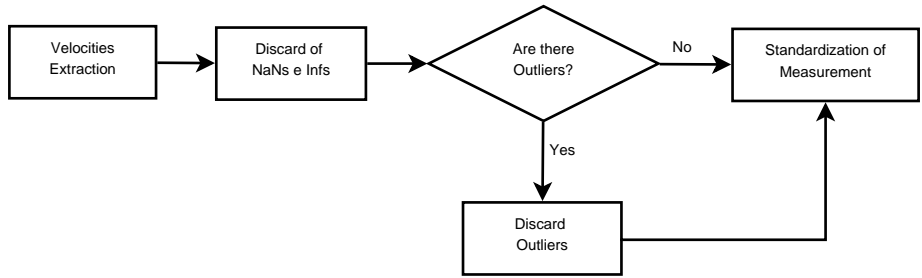} 
\caption{Methodology for data processing.} 
\label{fig:tratamento} 
\end{figure}

Data sets were previously processed in order to analyze the vehicle velocities through the causality Complexity-Entropy plane. 
The purpose is to extract the correct vehicle velocities information minimizing the presence of errors and outliers.
The methodology is presented in Figure~\ref{fig:tratamento}.

The first phase consists of extracting or calculating the velocities, if they are available in the data set or not, respectively. 
In the second phase \verb+NaN+s and \verb+Inf+s are discarded.
\verb+NaN+s mostly appear when there is a division by zero, for instance when the GPS registers two different positions occurring at the same time (instantaneous displacement).
\verb+Inf+s mostly appear when the GPS registers extremely large displacements in a very short time span.

Outliers are identified in the third phase, as those observations which are discrepant from the majority and present unreal velocities, for instance some velocities are bigger than \SI{200}{\kilo\meter\per\hour}.
If there are outliers the \textit{Discard outliers} is used to discard them, otherwise we pass directly to \textit{Standardization of Measures}.
In order to compare all data sets we include this last step to standardize the data in the same unit, in our case \si{\meter\per\second}.

The first data set used, \textit{Mobile Century}, includes velocities in \si{mi\per\hour}, and since neither wrong nor discrepant observations were found, only conversion to \si{\meter\per\second} was required.

The second data set used, \emph{Borl\"ange GPS}, required a thorough processing. 
Considering that all GPS samples are correct, in \textbf{phase one}, the velocities $v$ of each vehicle in each trip were calculated as the ratio of the vehicle displacement $\Delta s$ to the elapsed time $\Delta t$, i.e., 
$v = \Delta s/\Delta t$. 
The value of $\Delta s$ in \si{\meter} was computed as the shortest distance between two points (latitude, longitude)
using the \texttt{R} platform~\cite{R2012} which implements the Meeus~\cite{Meeus1999} method.

All files (\texttt{mobility}, \texttt{nodes} and \texttt{nodepos}) were used to calculate the velocities.
Table~\ref{tab:resumo} shows the time, displacements and velocities values of the example presented in Section~\ref{subsec:borlange}). 

\begin{table}[h] 
\centering 
\caption{Time, displacements and velocities values of some samples, \textit{Borl\"ange GPS} data set}
\label{tab:resumo} 
\begin{scriptsize} 
 \begin{tabular}{cccc} 
\toprule
file & time line & displacement (\si{\meter}) & velocity (\si{\meter\per\second}) \\ 
\midrule
1 &  14 & 229.53 & 16.39 \\ 
2 &  26 & 129.41 & 04.97 \\ 
3 &  19 & 271.84 & 14.30 \\ 
4 &  05   & 12.76 & 02.55 \\ 
5 &  54 & 306.45 & 05.67 \\ 
6 &  15 & 394.83 & 26.32 \\ 
7 &  00   & 300.00 & NaN \\ 
8 &  04   & 306.45 & 76.61 \\ 
\bottomrule
 \end{tabular} 
\end{scriptsize} 
\end{table} 

Wrong samples were removed in \textbf{phase two}.
Inconsistencies in the data set can be identified in Table~\ref{tab:resumo}. 
For instance, line $7$ would lead to a division by zero; all similar samples were discarded.

\begin{figure}[h] 
\centering 
\includegraphics[width=1\linewidth]{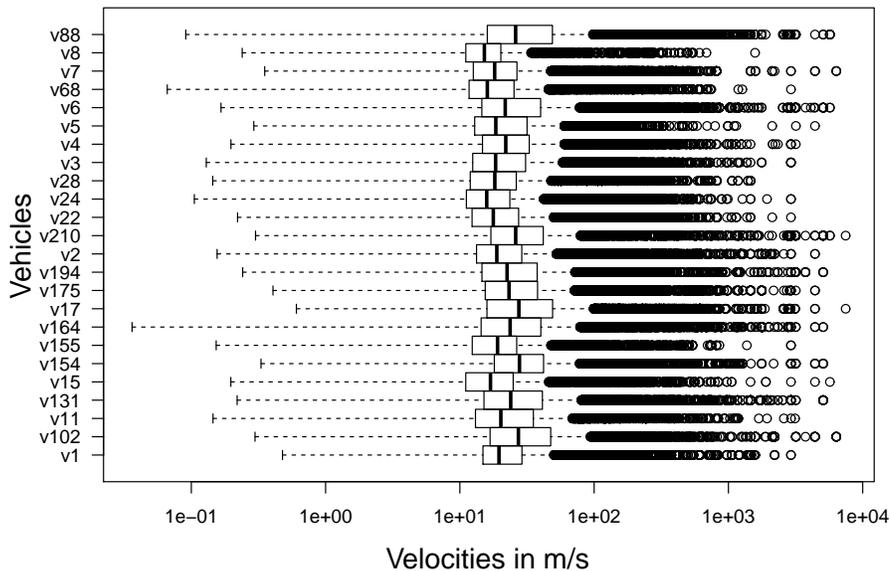} 
\caption{Boxplot of all individual velocities of all vehicles.} 
\label{fig:antes_tratamento} 
\end{figure}

Outliers were identified and discarded in \textbf{phase three} using a majority rule and common sense.
For instance, line $8$ on Table~\ref{tab:resumo}, we observe that the vehicle a velocity close to \SI{76}{\meter\per\second} (\SI{273}{\kilo\meter\per\hour}). 
To identify the outliers samples we use the Boxplot of all velocities of all vehicles (Figure~\ref{fig:antes_tratamento}). 

Figure~\ref{fig:antes_tratamento} shows the presence of velocities unfeasible for taxis (\SI{6,000}{\meter\per\second} or \SI{21,600}{\kilo\meter\per\hour}). 
To avoid an arbitrary discarding of velocities we analyze the data set in detail. 
Each trip is characterized by the mean velocity, and the lower and upper quartiles per trip were computed, and trips outside the interquartile range were discarded. 
The collection of all remaining trips is now analyzed, and observations above the upper quartile are considered outliers and discarded. 
The \textbf{fourth phase} was not necessary because the velocities are already in \si{\meter\per\second}.

\begin{figure}[h] 
\centering 
\includegraphics[width=1\linewidth]{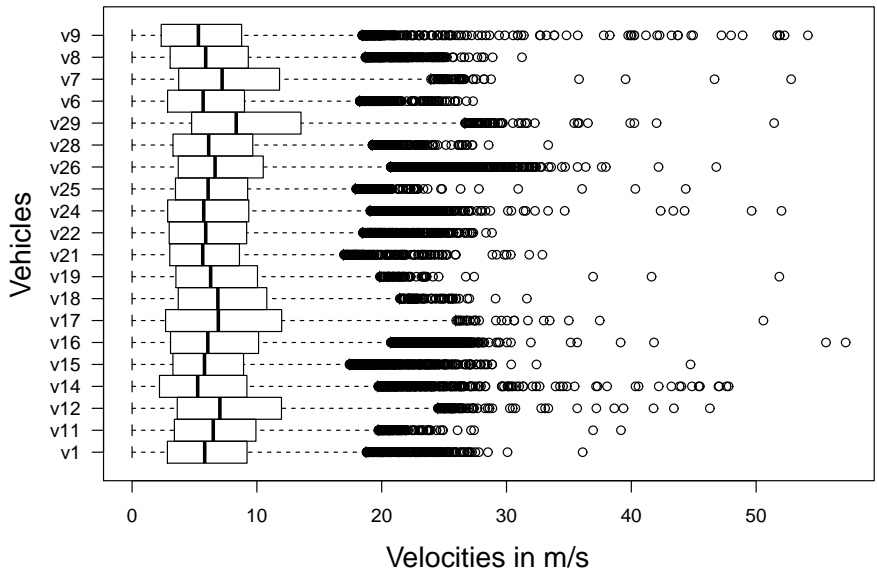} 
\caption{\emph{Boxplot} of velocities in Beijing database before the treatment.} 
\label{fig:boxplot_beijing_antes} 
\end{figure}

The last data set used, \emph{Beijing taxicabs}, needs some processing. 
In \textbf{phase one}, the data set ($30$ days log) had to be converted from binary to text. 
Many zero velocities were identified.
This occurs because taxis are not always in motion. 
Therefore, in analogy with \emph{Borl\"ange GPS} data set, in \textbf{phase two}, we organized the information by trips. 
Each trip starts when the velocity changes from zero to any other value and ends when the velocity returns to zero.
In \textbf{phase three} the velocities that are distant from the majority were identified. After that, they are discarded in the intermediate phase (\textit{Discard of outliers}). 
To identify the outliers samples we use the Boxplot of all velocities of all vehicles (Figure~\ref{fig:boxplot_beijing_antes}). 
All velocities above the upper quartile were discarded. 
\textbf{Phase four} was not necessary because the velocities already in \si{\meter\per\second}. 

\subsection{Velocities as time series}

Once velocities are computed and validated, the next step is to normalize all data sets as time series.
The time series are the concatenation of the velocities from all interpolated trips for each vehicle, discarding stopped vehicles. 
Observations must be equally sampled in order to facilitate the Information Theory quantifiers comparison, so Piecewise Cubic Hermite Interpolating Polynomials were used~\cite{Deboor1978}. 

\begin{figure}[h] 
\centering 
\subfigure[Interpolation]{\includegraphics[width=0.49\linewidth]{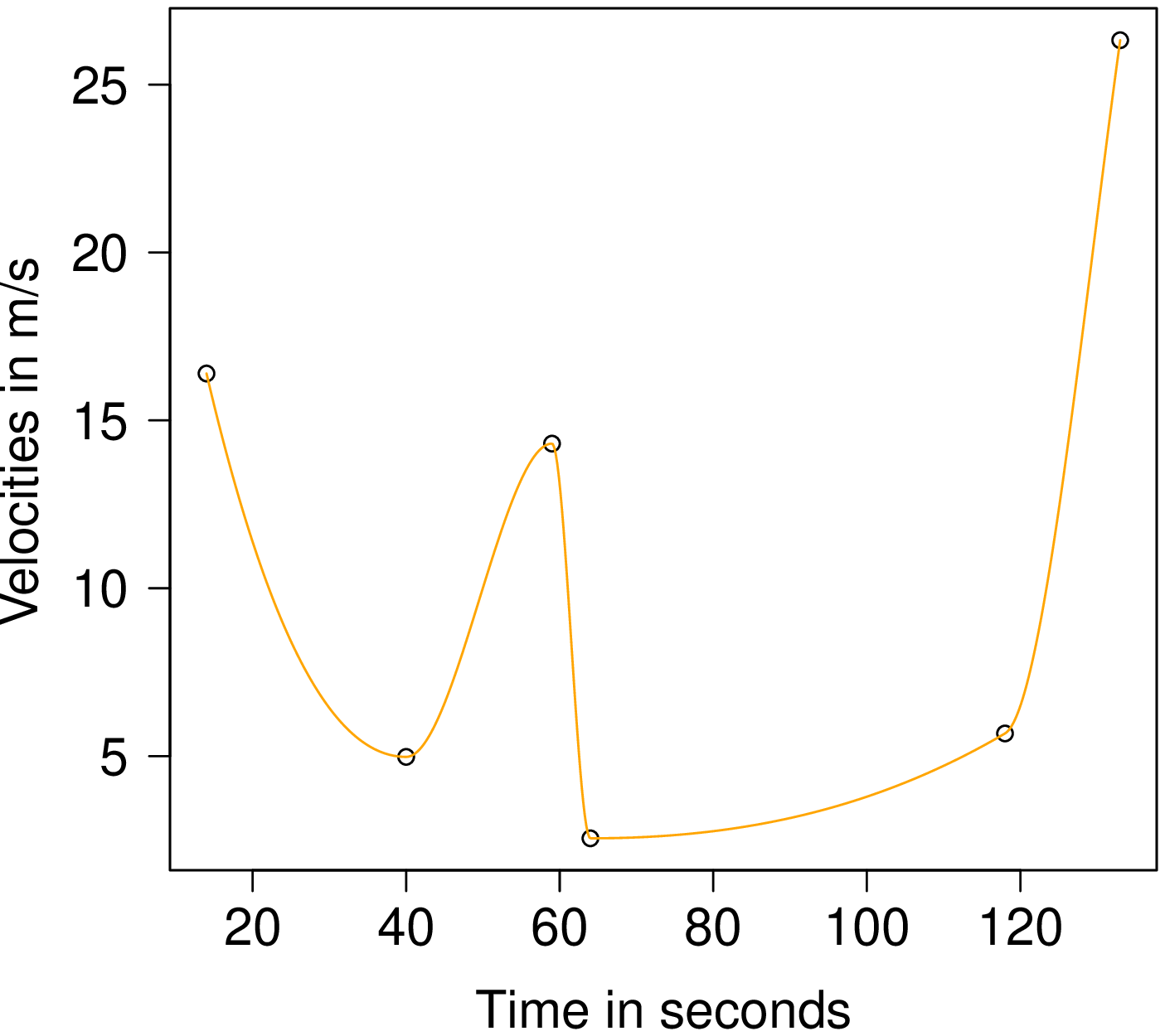}\label{fig:antes_interpolacao}}
\subfigure[New velocities]{\includegraphics[width=0.49\linewidth]{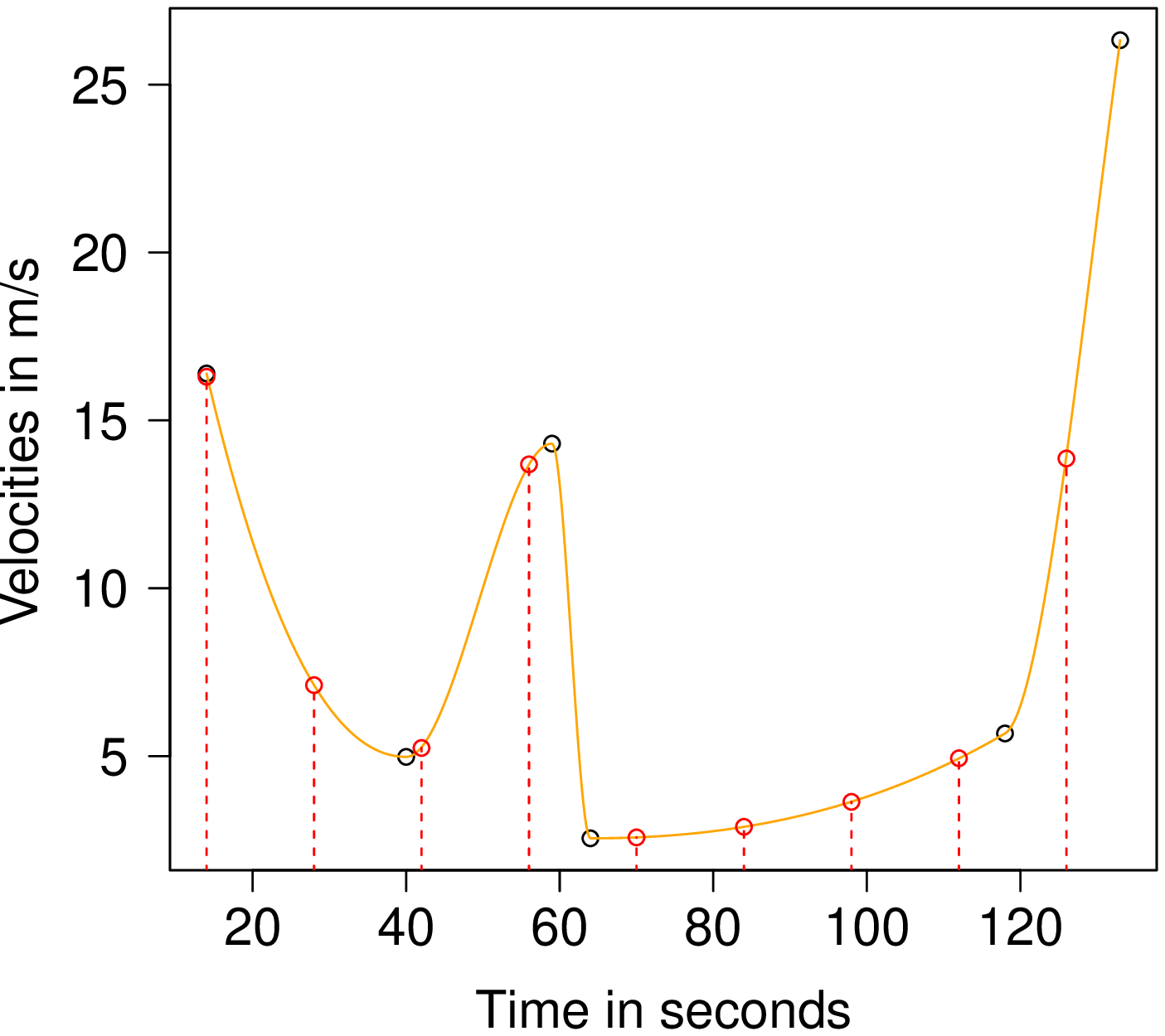}\label{fig:depois_interpolacao}} 
\caption{Interpolation process to sample of \textit{Borl\"ange GPS} data set.}  
\label{fig:interpolacao} 
\end{figure}

Figure~\ref{fig:antes_interpolacao} presents the interpolation of the velocities values from Table~\ref{tab:resumo} after processing. 
The new velocities were obtained sampling at a constant intervals $T_{S}=$ \SI{14}{\second}, as shown in Figure~\ref{fig:depois_interpolacao}.
This interval value was based on the average time interval considering all valid trips.

\begin{figure}[h] 
\centering 
\subfigure[\emph{Mobile Century}]{\includegraphics[width=0.49\linewidth]{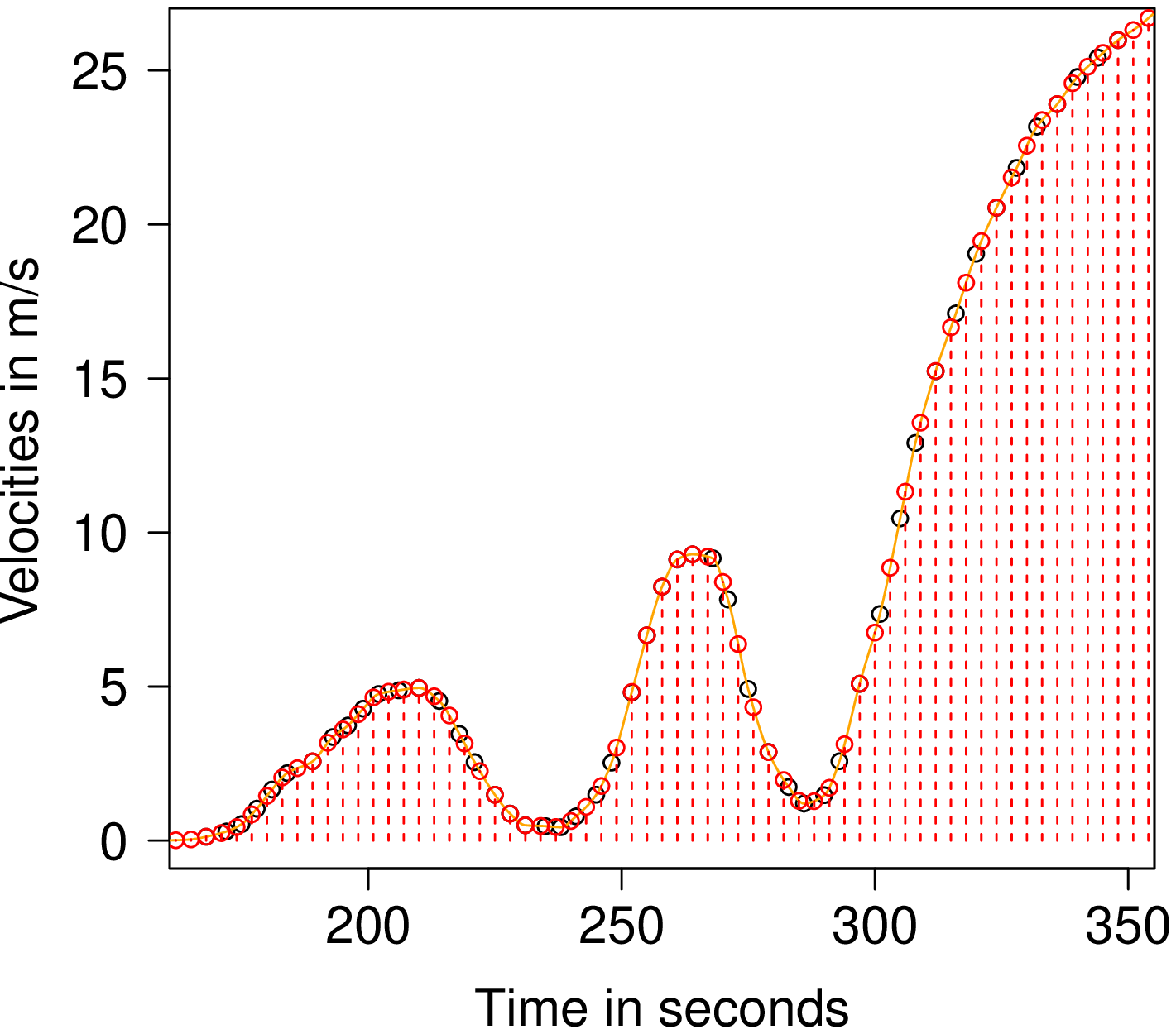}\label{fig:depois_interpolacao_m}}
\subfigure[\textit{Beijing taxicabs}]{\includegraphics[width=0.49\linewidth]{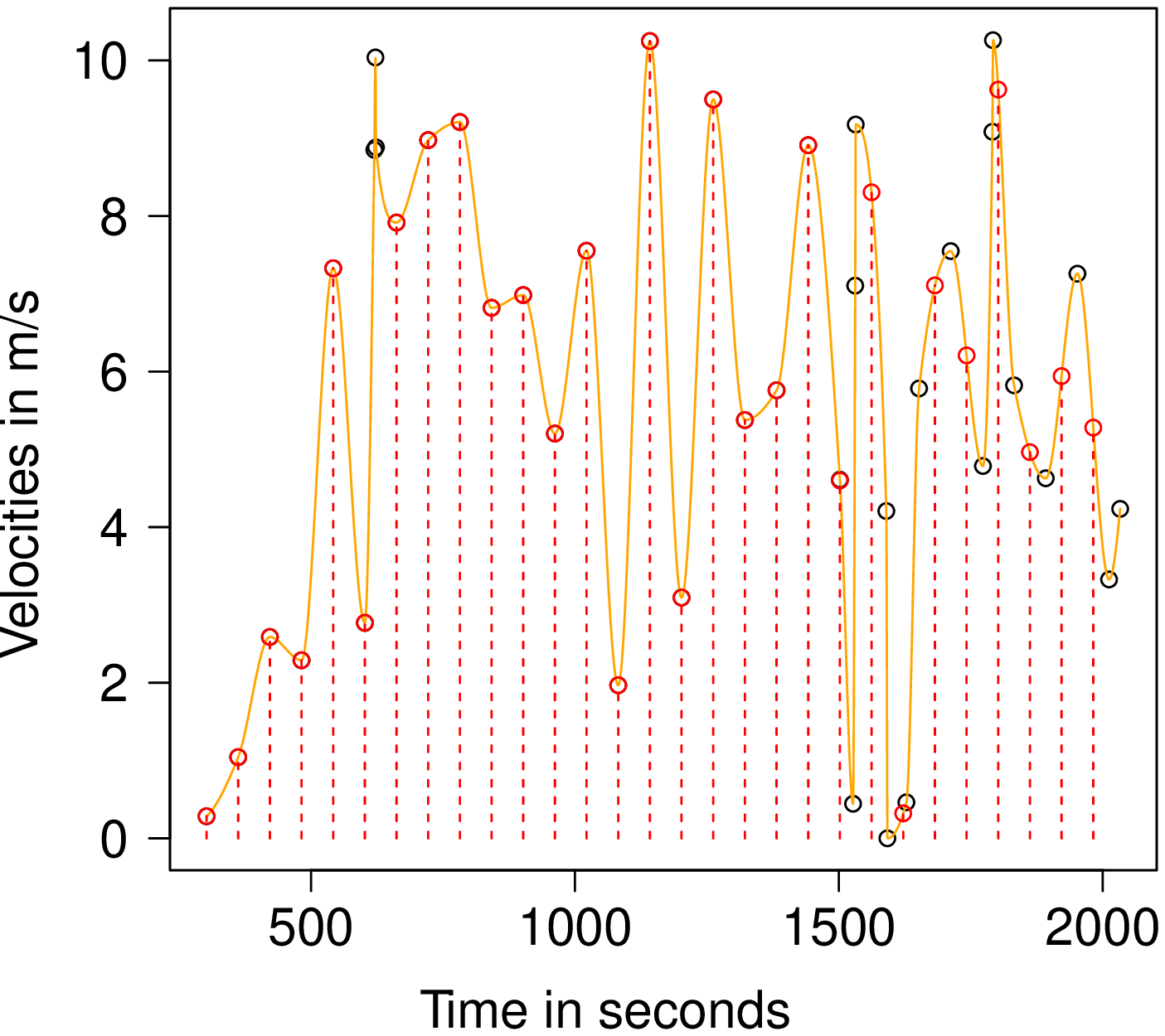}\label{fig:depois_interpolacao_b1}}
\caption{Result of interpolation process to samples of \emph{Mobile Century} and \textit{Beijing taxicabs} data sets.} 
\label{fig:interpolacao_m} 
\end{figure}

The same interpolation processing was used to \emph{Mobile Century} and \textit{Beijing taxicabs} data set. 
We use, respectively, intervals $T_{S}=$ \SI{3}{\second} and $T_{S}=$ \SI{60}{\second}.
Figure~\ref{fig:interpolacao_m} presents the velocities after interpolation.
 
The initial time series sizes are $1,795$, $1,191$, and $3,463$ for the \textit{Borl\"ange GPS}, \textit{Mobile Century} and \textit{Beijing taxicabs} data sets, respectively. 
After interpolation, the resulting time series sizes are $19,456$, $8,200$, and $14,461$.
Different time series sizes are considered because each real data set presented different number of samples, so we consider all of them without compromising the results and analysis presented.
In all cases, the size of time series $M$ is much greater than the embedding dimensions (pattern length) $D = 4$, respecting the constraint of the method wherein $(M \gg D!)$, as presented in Section~\ref{sec:methods}.

\section{Stochastic $f^{-k}$ power spectrum data}
\label{sec:stochastic_data}

\begin{figure}[h] 
\centering 
\subfigure[White ($k = 0$)]{\includegraphics[width=0.45\linewidth, trim=50 50 50 50, clip=true]{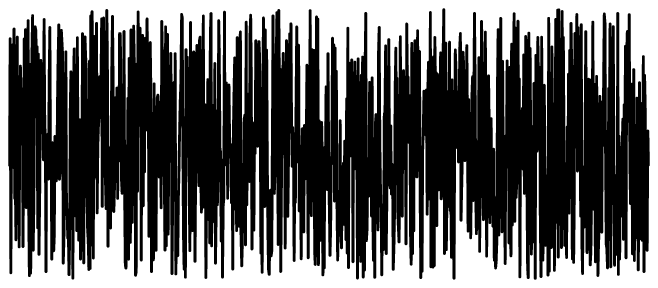}\label{fig:ruido_branco}} \qquad
\subfigure[Pink ($k = 1$)]{\includegraphics[width=0.45\linewidth, trim=50 50 50 50, clip=true]{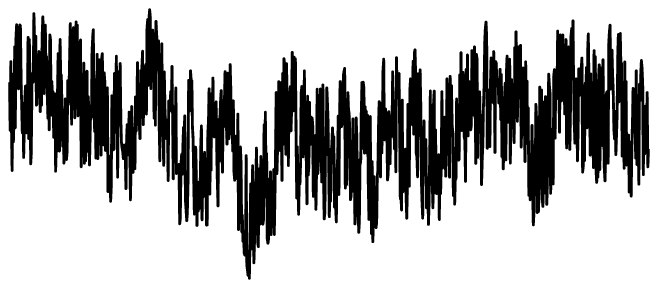}\label{fig:ruido_rosa}} 
\subfigure[Brown ($k = 2$)]{\includegraphics[width=0.45\linewidth, trim=50 50 50 50, clip=true]{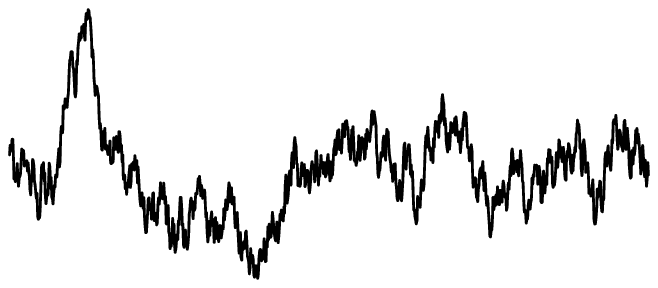}\label{fig:ruido_marrom}} \qquad
\subfigure[Black ($k > 2$)]{\includegraphics[width=0.45\linewidth, trim=50 50 50 50, clip=true]{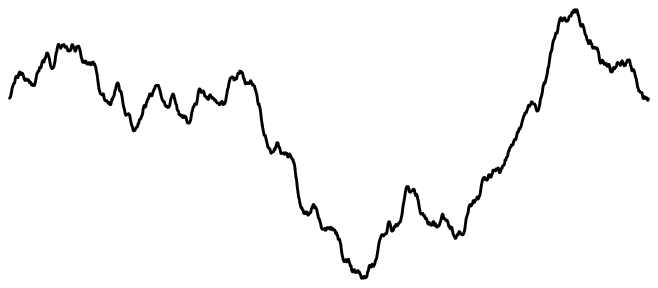}\label{fig:ruido_preto}} 
\caption{Examples of colored noises.} 
\label{fig:ruidos} 
\end{figure}

Schroeder~\cite{Schroeder2009} ranked noises, observing their Power Spectral densities and assuming that they can be described by a frequency function $f^{-k}$. 
He identified four groups: white, pink, brown, and black noise, as depicted in Figure~\ref{fig:ruidos}.

\begin{itemize}
\item In the white noise ($k = 0$, Figure~\ref{fig:ruido_branco}), the power spectrum is independent of the frequency, i.e., is constant. 
Its name comes from the analogy with white light, which possess all frequencies of the visible spectrum. 

\item Pink noise ($k = 1$, Figure~\ref{fig:ruido_rosa}) is known as $1/f$ noise and can be found in various physical processes.
It is commonly used in music and arts, for example, as input in research on hearing and acoustics
~\cite{Schottky1926}. 
It is similar to natural noises, as the rain or waterfall sounds~\cite{Voss1979}.

\item The brown noise ($k = 2$, Figure~\ref{fig:ruido_marrom}) can be generated by the integration of white noise over time. 
Example of this noise is the Random Way, i.e., a random choice is taken to determine the direction of a movement~\cite{Lawler2010}.

\item The black noise ($k>2$, Figure~\ref{fig:ruido_preto}) is known as silent noise, and can be used to model natural disasters such as floods, droughts, electrical failures etc. 
Negligible powers predominate in its frequency spectrum, except for a few narrow bands.
\end{itemize}

In order to compare VANETs data with the stochastic ones, we generate the noises with $f^{-k}$ as in Ref.~\cite{Rosso2007a}: 
(a)~A function of the statistical software \texttt{R}~\cite{R2012} is used to produce pseudo random numbers in the interval $[-0.5,0.5]$ with an (i)~almost flat power spectrum, (ii)~uniform probability distribution function, and (iii)~zero mean value. 
(b)~Then, the fast Fourier transform (FFT) $y_k^1$ is obtained and multiplied by $f^{-k/2}$, yielding $y_k^2$. 
(c)~Now, $y_k^2$ is symmetrized so as to obtain a real function and then the pertinent inverse FFT $x_j$ is obtained, after discarding the small imaginary components produced by numerical approximations. 
The ensuing time series $x_j$ has the desired power spectrum.

\section{Results and discussions}
\label{sec:results}

In this proposal, we associate a probability distribution function to time series by Bandt-Pompe symbolization (see Section~\ref{sec:methods}). 
This method considers a temporal causality, by comparing the current values with their neighbours in the time series.
Bandt and Pompe~\cite{Bandt2002} suggest to use an embedding dimensions between 3 and 7.
Thus, we use the number of embedded dimensions ($D$) equal to $4$, and the embedding delay time equal to 1. 

\begin{figure*}[t] 
\centering 
\subfigure[\emph{Mobile Century}]{\includegraphics[width=0.32\linewidth]{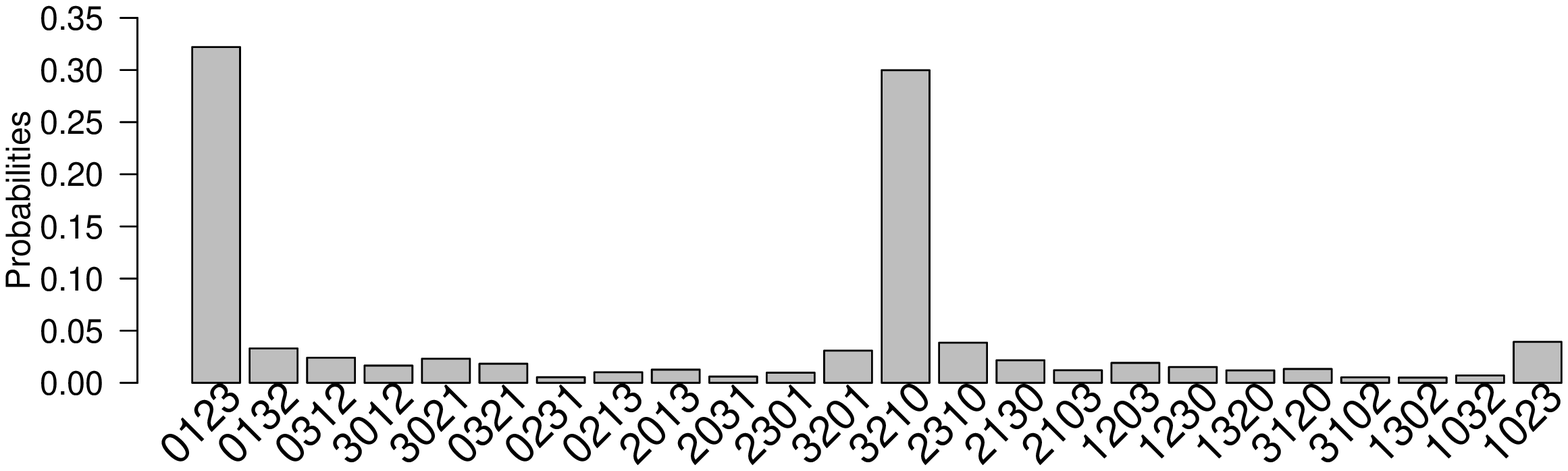}\label{fig:prob_mobile}}
\subfigure[\textit{Borl\"ange GPS}]{\includegraphics[width=0.32\linewidth]{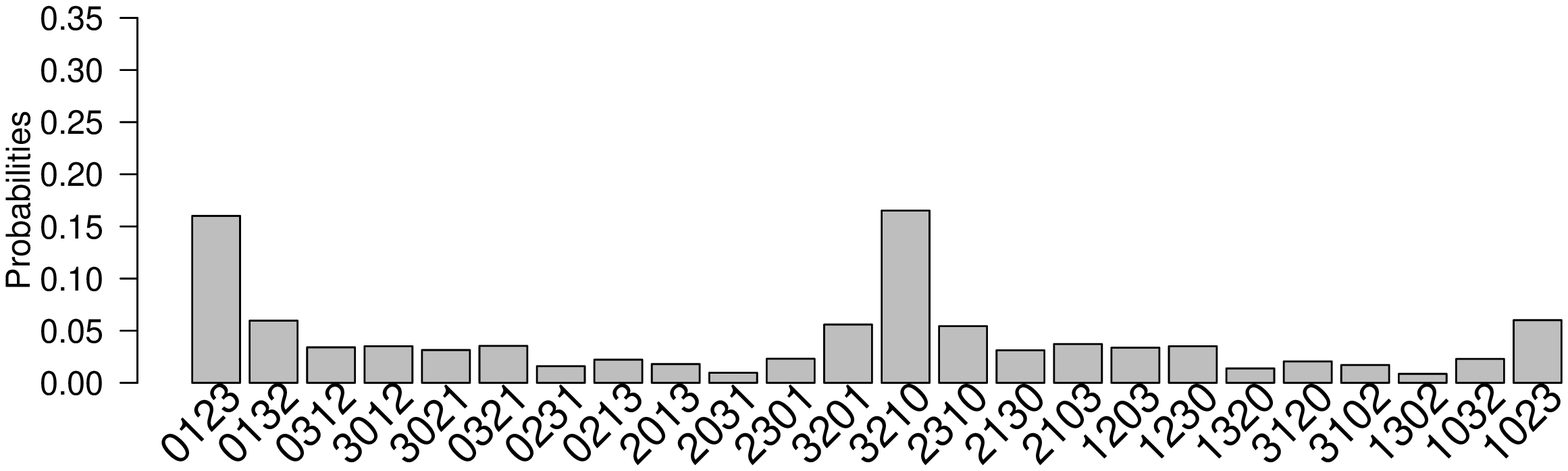}\label{fig:prob_borlange}}
\subfigure[\textit{Beijing taxicabs}]{\includegraphics[width=0.32\linewidth]{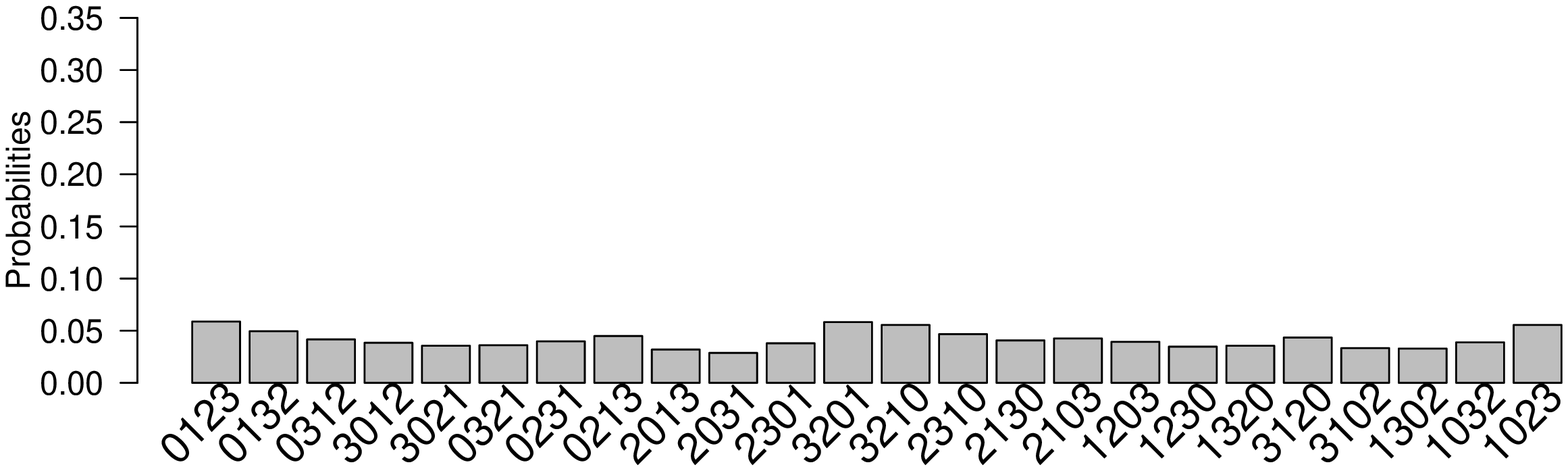}\label{fig:prob_beijing}} 
\caption{Examples of probability distribution functions from the patterns of the three data sets.}
\label{fig:bandtepompe} 
\end{figure*}

The Figure~\ref{fig:bandtepompe} presents the probability distribution functions of patterns from \emph{Mobile Century}, \textit{Borl\"ange GPS} and \textit{Beijing taxicabs} data sets. 
For instance, the pattern 0123 means that the vehicle velocity in 4 samples increased, for instance, $t_0 \rightarrow v=\SI{5}{\meter\per\second}$, $t_1 \rightarrow v=\SI{10}{\meter\per\second}$, $t_2 \rightarrow v=\SI{15}{\meter\per\second}$, and $t_3 \rightarrow v=\SI{20}{\meter\per\second}$. 
Figure~\ref{fig:prob_borlange} shows that the probability of this pattern is close to $15\%$, in the \textit{Borl\"ange GPS} data set.

Observing the probabilities functions in Figure~\ref{fig:bandtepompe}
it is possible to identify differences among the data sets. 
Specifically, the Figure~\ref{fig:prob_mobile} with \emph{Mobile Century} data set presents high values associated to patterns 0123 and 3210.
This occurs because the samples were collected in a highway, so the velocities do not present erratic variations as the vehicles always increase or decrease their velocities.
In the same way, but in a smaller proportion, Figure~\ref{fig:prob_borlange} shows that the \textit{Borl\"ange GPS} data set also presents high values in patterns 0123 and 3210. 
In this case, this occurs because the samples were collected in a region with highways and city roads.
Finally, Figure~\ref{fig:prob_beijing}, \textit{Beijing taxicabs} data set, presents a uniform variation in the probabilities of velocities patterns.
This occurs because the samples were collected in a central business region with high traffic jams and consequently more velocities variation.
Traffic behavior can be further analyzed using these distributions to compute the
permutation Normalized Shannon Entropy and permutation Statistical Complexity. 

\begin{figure}[h] 
\centering 
\subfigure[View of all data sets.]{\includegraphics[width=.45\linewidth]{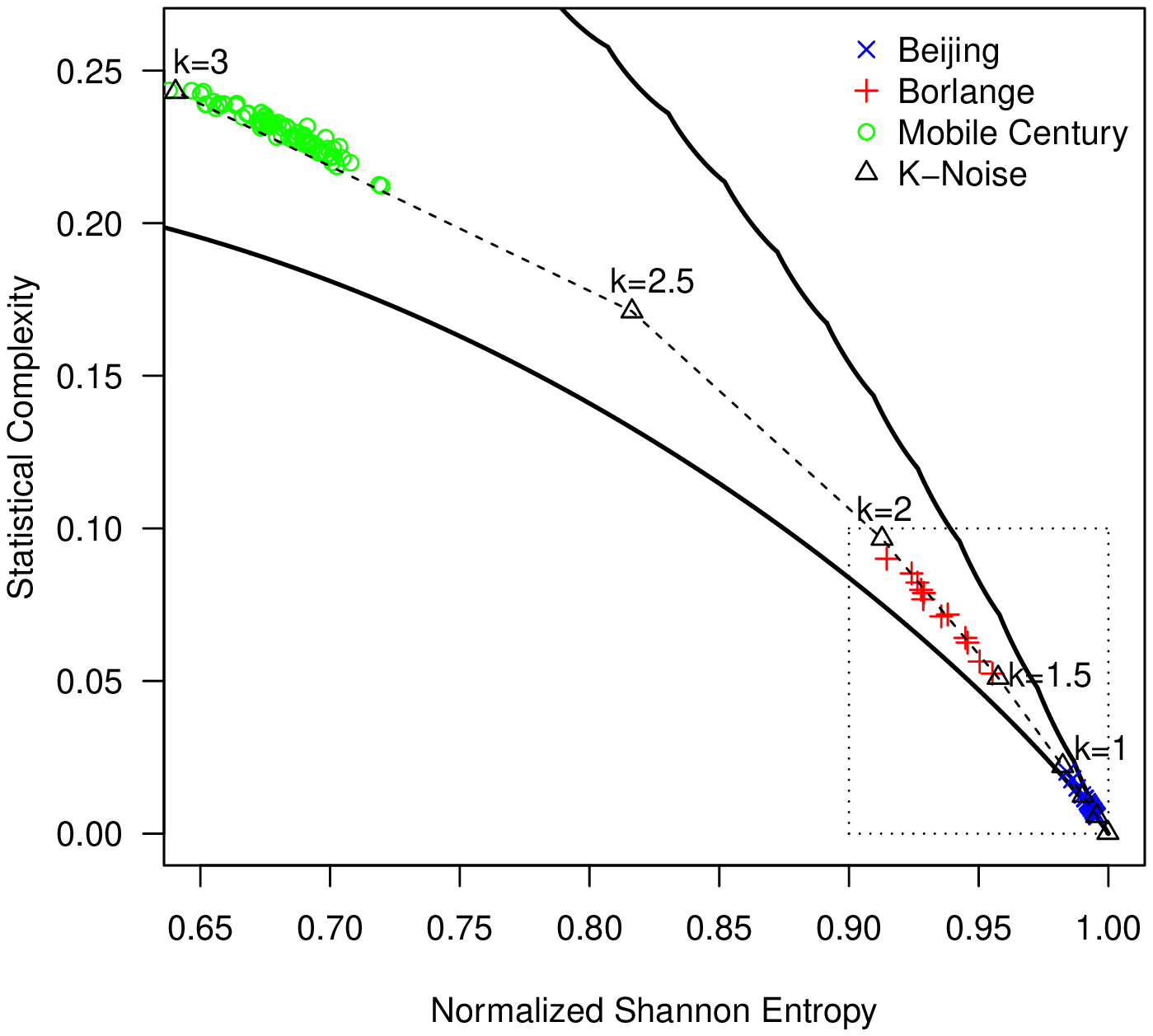}\label{fig:resultado1}}
\subfigure[Zoom of Borl\"ange and Beijing data sets.]{\includegraphics[width=.45\linewidth]{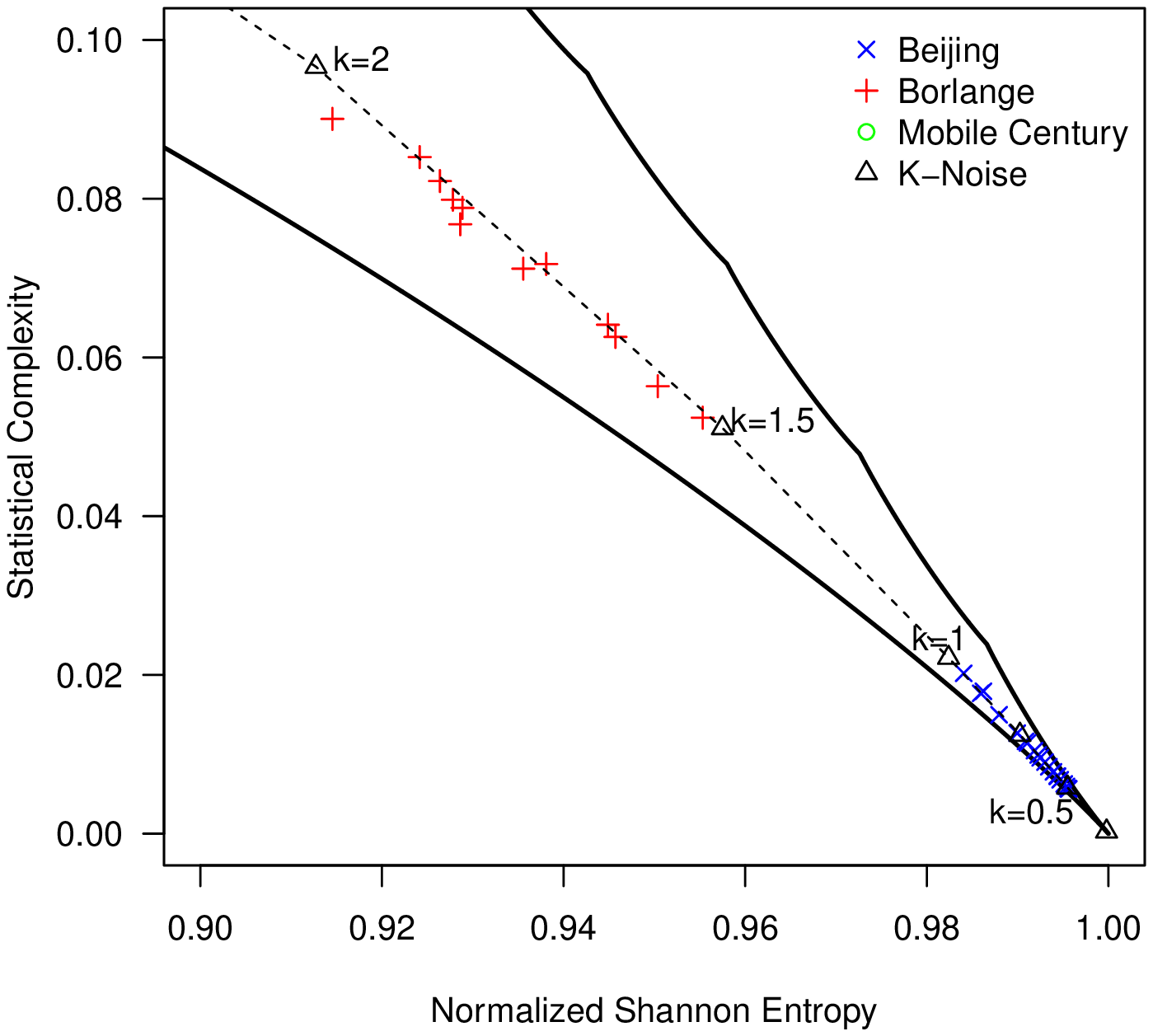}\label{fig:resultado2}}
\caption{Complexity-Entropy plane.}
\label{fig:resultado} 
\end{figure}

Figure~\ref{fig:resultado} shows the causality Complexity-Entropy plane with embedded dimensions $D = 4$.  
The information from the three vehicles data sets (77 circle points from \emph{Mobile Century} ), 12 cross points from \textit{Borl\"ange GPS}, and 20 times points from \emph{Beijing taxicabs})
is presented along with
colored noise $f^{-k}$, $k=\{0, 0.5, 1, 1.5, 2, 2.5, 3\}$ (triangle points).
The dashed line was used to clarify the data visualization.

It is noticeable that vehicles cluster, and that each data set is closer to a different colored noise.
This is an important result because all inference about colored noise can be used to traffic behavior.
Specifically, the \emph{Mobile Century} vehicles (Figure~\ref{fig:resultado1}) range in the interval $2.5 \leq k \leq 3$, that is, between the Brown and Black noises. 
In this case, we can infer that the velocities in highways are highly correlated.
Additionally, we can associate this range with a persistent Fractional Brownian Motion ($2 < k < 3$), i.e., a long memory processes~\cite{Rosso2007a}.

Figure~\ref{fig:resultado2} is a zoom over the \emph{Borl\"ange GPS} observations, which lie close to the Brown Noise.
We observe that there are some regions with high concentration of vehicles and others free of them. 
In this kind of scenario, the velocities are less correlated when compared with the previous one. 
This behavior can be associated with a antipersistent Fractional Brownian Motion ($0.5 < k < 2$), i.e., a short memory processes~\cite{Rosso2007a}.

Finally, Figure~\ref{fig:resultado2} shows a zoom of the \emph{Beijing taxicabs} observations, which lie close to the Pink Noise.
We can identify that the velocities in a busy city downtown, i.e., with heavy traffic, are highly variable and poorly correlated.
This behavior can be associated with a antipersistent Fractional Brownian Motion or with a Fractional Gaussian Noise ($0.5 < k < 2$)~\cite{Rosso2007a}. 
 
This is an important result because, in general, any traffic behavior is simulated as random walk, i.e., a stochastic process (white noise $k=0$). 
However, as presented, this assumption is far from what is observed in practice.
Thus, traffic behavior simulation should be simulated with correlated noise $f^{-k}$ with $0.5 < k < 3$, and this study provides different values for different plausible situations.

Based on the discussion of Section~\ref{sec:time_vs_plane},
we exemplify that behavior for \textit{Mobile Century} data set when
the time sampling $T_{S}=$ \SI{3}{\second} was increased three times
($T_{S}'=$ \SI{9}{\second}).
For $T_{S}=$ \SI{3}{\second} the points are located in the plane
in the range $0.64 \leq {\mathcal H} \leq 0.73$ and $0.21 \leq
{\mathcal C} \leq 0.24$, which is compatible with $k$-noise $\simeq 2.87$.
In the case of $T_{S}'=$ \SI{9}{\second} we have $0.75 \leq {\mathcal
H} \leq 0.85$ and $0.15 \leq {\mathcal C} \leq 0.20$, which is compatible with
$k$-noise $\simeq 2.68$.
The same analysis was performed with the \emph{Borl\"ange GPS} data set.
The original time sampling $T_{S}=$ \SI{14}{\second} was increased to $T_{S}'=$ \SI{30}{\second}. 
For $T_{S}=$ \SI{14}{\second} the points are located in the plane in the range $0.85 \leq {\mathcal H} \leq 0.95$ and $0.05 \leq
{\mathcal C} \leq 0.12$, which is compatible with $k$-noise $\simeq 1.7$.
In the case of $T_{S}'=$ \SI{30}{\second} we have $0.93 \leq {\mathcal
H} \leq 0.99$ and $0.01 \leq {\mathcal C} \leq 0.06$, which is compatible with
$k$-noise $\simeq 1.3$.
These examples illustrate how changes in the sampling rate may affect the observed dynamics.

The  Borl\"ange GPS data correlation is lost when sampled at \SI{60}{\second}.
This indicates that too much data is lost in this case, and that such
timescale is too far from the optimal.

\section{Conclusion} 
\label{sec:conclusions}

This work proposed a study to characterize the velocities behavior.
This characterization is needed for a better understanding of the underlying dynamics governing this process, and to improve the design of VANETs applications.
This characterization, differently from previous works, identifies the underlying global dynamics, using velocities as the only source of information.
We used the Bandt-Pompe methodology to assign a probability distribution function (PDF) to time series that describe the vehicles velocities. 
The characterization was performed using two Information Theory concepts: Entropy and Complexity. 

The data from vehicles were extracted from three data sets: \emph{Mobile Century}, \textit{Borl\"ange GPS} and \textit{Beijing taxicabs}. 
These data sets were subjected to a number of transformations and imputations in order to have comparable data as free of contaminated observations as possible.
As we have no control over the sampling rates, we rely on using the same interpolation technique on all data sets.
This sampling allows the characterization of the underlying dynamics, and they agree with the kind of  process induced by the physical topology and characteristics of each data set.
We, thus, consider this time as a good approximation to the ``optimal'' time-sample in the sense that it captures the velocity time series representative of the correct dynamics associated with the vehicle velocities set under study. 
Clearly, different velocities data sets will have different ``optimal'' time-sampling characteristics, 
lead to rather stable points in the causality entropy-complexity plane, as they are characteristic of the available 
information about the underlying dynamics which generates the corresponding time series under analysis,
and, therefore, the points can be compared.

We plotted the permutation entropy and permutation statistical complexity from Bandt-Pompe probability 
distributions of vehicles velocities. 
We observe that the behavior of the velocities is similar to colored noise with $k$ ranging in the interval $0.5 \leq k \leq 3$. 
Therefore we can infer that the global behavior of velocities is compatible to this noise. 
This is an important result because, in general, any traffic behavior is simulated as random walk, i.e., a stochastic process (white noise $k=0$).
This hypothesis has been rendered inadequate, in real scenarios, by this study.

The VANETs time series here analyzed allow to conclude that the data
are generated by complex dynamical systems with different global
behavior at distinct time scales;
In particular, the correlation is lost in the Borl\"ange GPS data set
when the sampling rate is increased to \SI{60}{\second}.
This result is a clear indication that the sampling rate plays a
central role in the modeling of vehicles velocities: as colored noise
at short timescales, and as white noise at long timescales.
As white noise has no information, this result highlights the
importance of sampling the data at optimal rates, which are typically
high.
Otherwise, the underlying behavioral structure will be lost and the
system will not be characterized or identified.

Our next step consists in applying the proposed characterization to other data sets. 
We intend to apply this methodology to characterize the vehicle positions instead velocities.
Another work direction is the use of colored noise to improve the trajectories prediction systems. 
Our approach will be used to analyze different cycles along the day, as in~\cite{Liao2012}, provided more data are available.

\section*{Acknowledgements}

We acknowledge support from the Brazilian research agency CNPq and the Research Foundation of the State of Alagoas (FAPEAL)
O. A. Rosso gratefully acknowledges support from CONICET, Argentina.

\section*{Author Contributions}

All authors contributed equally to the paper.

\bibliographystyle{elsarticle-num}
\bibliography{reference-minimal}

\end{document}